\documentclass[aps,supepscriptaddress,showpacs,nofootinbib]{revtex4-1}
\usepackage{bm}
\usepackage{graphicx}
\usepackage{tikz}
\usepackage{chemarrow}

\begin{document}

\title{Chiral autocatalysis: reaction noise, micro-reversibility and chiral inhibition in mirror symmetry breaking}

\author{Michael Stich$^1$, Josep M. Rib\'{o}$^2$ and David Hochberg$^3$}
\email{hochbergd@cab.inta-csic.es} \affiliation{$^1$Non-linearity
and Complexity Research Group, School of Engineering and Applied
Science, Aston University, B4 7ET Birmingham, UK}
\affiliation{$^2$Department of Organic Chemistry, Institute of
Cosmos Science (IEEC-UB), University of Barcelona, Barcelona, Spain}
\affiliation{$^3$Department of Molecular Evolution, Centro de
Astrobiolog\'{\i}a (CSIC-INTA), Carretera Ajalvir Kil\'{o}metro 4,
28850 Torrej\'{o}n de Ardoz, Madrid, Spain}

\begin{abstract}
Applying the constraints dictated by the principle of detailed
balance, we analyze a recent proposal for spontaneous mirror
symmetry breaking (SMSB) based on enantioselective autocatalysis
coupled to a linear decay of the enantiomers and in the presence of
reaction noise. We find the racemic state is the final stable
outcome for both deterministic as well as for stochastic dynamics,
and for both well-mixed and small spatially-coupled systems. The
racemic outcome results even when the autocatalytic cycles are
driven irreversibly by external reagents, in manifestly
non-equilibrium conditions. Our findings suggest that first-order
autocatalysis coupled to reactions involving \textit{non-linear}
heterochiral dynamics is a necessary pre-condition for any mechanism
purporting to lead to molecular homochirality.
\end{abstract}

\date{\today}

\maketitle

\section{\label{sec:intro} Introduction}

The observed bias in biopolymers composed from homochiral L-amino
acids and D-sugars towards a single handedness or chirality is a
remarkable feature of biological chemistry. Nowadays, there is a
firm consensus that the homochirality of biological compounds is a
condition associated to life that probably emerged in the prebiotic
phase of evolution through processes of spontaneous mirror symmetry
breaking (SMSB) \cite{Cintas}. This could have proceeded by
incorporating steps of increasing complexity thus leading to
chemical systems and enantioselective chemical networks
\cite{AG,Guijarro}. Theoretical proposals for the emergence of
homochirality in abiotic chemical evolution, are based either on
deterministic or on chance events \cite{AG,Guijarro}. However, the
current state of knowledge strongly suggests that the emergence of
chirality must be based on reactions leading to spontaneous mirror
symmetry breaking. SMSB are transformations yielding chiral outcomes
as non-thermodynamic final stable states, and in the absence of any
chiral polarization or external chiral physical forces \cite{smsb}.
This is provided by enantioselective autocatalysis, but not by the
simple linear asymmetric induction reactions \cite{AAS-review} on
which past discussions on deterministic or chance phenomena were
based for the justification of biological homochirality. Systems
capable of SMSB lead to a stochastic distribution of final chiral
signs between successive experiments. Nowadays this deterministic
versus chance debate is restricted to more specific scenarios
\cite{Cintas,AAS-review, Lahav2005}. The SMSB abiotic scenario for
the emergence of single homochirality in the biological world
implies that single asymmetry emerges provided a small chiral
fluctuation with respect to the idealized racemic state can be
amplified \cite{amplify} to a state capable of biotic evolution.
Relevant features common to such systems are that they take into
account the small fluctuations about the racemic state and that they
display \textit{non-linear} kinetic effects. These stochastic
scenarios are theoretically well understood on general grounds
\cite{Konde1985,Avetisov1987} and equally important, are
experimentally feasible in the laboratory \cite{Lahav2005,Soai,
Viedma}.

On the theoretical side, for describing systems capable of SMSB, the
Frank model has been widely invoked to justify the emergence of
biological homochirality \cite{Guijarro,Cintas}. The original model
\cite{Frank} consists of an irreversible enantioselective
autocatalysis Eq.(\ref{Frank1953}) and an irreversible mutual
inhibition reaction Eq.(\ref{mutual}) between the product
enantiomers, in an open-flow system:
\begin{eqnarray}\label{Frank1953}
A + L \stackrel{k_a}{\longrightarrow} L + L, &&\qquad  A + D
\stackrel{k_a}{\longrightarrow} D + D,\\ \label{mutual}
 L + D && \stackrel{k_2}{\longrightarrow} P.
\end{eqnarray}
This model has been extensively studied, allowing for reversible
transformations and in diverse open-system scenarios \cite{Plasson}.
The significance of the mutual inhibition step Eq. (\ref{mutual}) is
that it makes SMSB possible for first-order enantioselective
autocatalysis, such as that of Eq. (\ref{Frank1953}). Although
enantioselective autocatalyses of quadratic and higher orders may by
themselves lead theoretically to SMSB, they correspond to reactions
of unrealistically high molecularity. For example, biological
replicators of interest for enantioselective autocatalytic
transformations, have their mechanisms composed by consecutive
bi-molecular reactions. This means that, by themselves, these
replicators \textit{cannot amplify} the initial enantiomeric excess
($ee$). However, the coupling to a mutual inhibition reaction
between the enantiomers can enable SMSB for some reaction and system
parameters. Therefore, the chemically significant scenarios for the
study of the emergence of chirality are those consisting of simple
linear enantioselective autocatalyses coupled to reaction networks
that include a mutual inhibition step.

Originally proposed as an alternative to the Frank model, the
limited enantioselectivity (LES) model is composed of entirely
reversible steps: an enantioselective autocatalytic reaction
Eq.(\ref{LESscheme}), a limited enantioselective autocatalysis
Eq.(\ref{limited}), and direct production of enantiomers from an
achiral precursor Eq. (\ref{direct}) \cite{AG}:
\begin{eqnarray}\label{LESscheme}
A + L  \autorightleftharpoons{$k_a$}{$k_{-a}$} L + L, &&\qquad A + D
\autorightleftharpoons{$k_a$}{$k_{-a}$} D + D, \\ \label{limited} A
+ L \autorightleftharpoons{$k_e$}{$k_{-e}$} L + D, &&\qquad A + D
\autorightleftharpoons{$k_e$}{$k_{-e}$}
D + L, \\
\label{direct} A \autorightleftharpoons{$k_n$}{$k_{d}$} L, &&\qquad
A \autorightleftharpoons{$k_n$}{$k_{d}$} D.
\end{eqnarray}
Note that the \textit{inverse} reaction in Eq.(\ref{limited}) with
rate $k_{-e}$ provides the necessary chiral inhibition step, thus
effectively replacing Frank's mutual inhibition Eq. (\ref{mutual})
leading to the inert product $P$. The dynamic stability properties
of racemic and chiral states in fully reversible versions of Frank
and in LES, including hetero- and homo-dimerization, in both open
and closed systems, are reported in detail in \cite{RH}.

Typically, rate equation theory (in the mean field approximation) is
used to cast chemical reaction schemes in terms of coupled
differential equations for the temporal evolution of the
concentrations of the chemical species involved. In this
deterministic kinetic dynamics, \textit{initial conditions} must be
taken to simulate the inherent statistical chiral fluctuations about
the ideal racemic composition \cite{Mills, Mislow}. In contrast,
real chemical reactions are inherently stochastic in nature: the
reagents in solution (or attached on surfaces) must encounter each
other before they can react, and the probability per unit time for
the reaction to occur is related to the corresponding reaction rate
constant, which depends on the temperature. The molecular nature of
chemical reagents and reactions gives rise to the concept of
intrinsic \textit{reaction noise}, and is typically multiplicative
\cite{vanKampen}. Despite the fact that stochastic and deterministic
kinetics must coincide in the macroscopic limit (i.e., for large
numbers of molecules), stochastic methods can be used to address the
question of whether such internal noise affects the final outcome of
the underlying reaction, and in what way it might do so. The answer
to this question depends on the specific process studied. Thus, for
example, reaction noise explains the anomalous scaling in reactions
undergoing dynamic phase transitions from active to absorbing states
\cite{Tauber}.

The influence that reaction noise may have in schemes purporting to
lead to SMSB is readily investigated. In the case of the Frank model
with \textit{reversible} autocatalysis (defined by the pair of
Eqs.(\ref{LESscheme},\ref{mutual})), reaction noise induces complete
chiral amplification starting from ideally racemic initial
conditions in spatially extended domains, but this result requires
the parameter $g \equiv \frac{k_{-a}}{k_2}$, controlling the
symmetry breaking transition, to satisfy the same condition as for
the deterministic model, namely $0 \leq g < 1$ \cite{HZcpl,HZpre}.
That is, the \textit{final outcome} of the reaction scheme is
governed solely by this simple condition, irrespective of whether
internal noise is included in the temporal evolution or not. On the
other hand, neither chiral bias nor external physical chiral
polarizations need be invoked to achieve homochirality when reaction
noise is included \cite{HZcpl}. Hence, the initial and subsequent
chiral fluctuations intrinsic to the system \cite{Mills, Mislow} are
included automatically.

Stochastic methods are necessary to describe kinetic dynamics in the
case of small volumes and/or small numbers of reacting molecules
\cite{Gillespie2007,Puchalka}, as is the case, for example, in
compartmentalized cellular processes \cite{ErdiLente}. Therefore,
the differences in the evolution of the $ee$ between deterministic
and stochastic kinetics should provide better insights regarding
asymmetric inductions and SMSB processes in living systems. In this
respect, internal noise has been considered recently in a
closed-mass model which results from taking strictly irreversible
enantioselective autocatalysis, Eq.(\ref{Frank1953}) together with
the direct production and decay of the enantiomers,
Eq.(\ref{direct}) \cite{Goldenfeld}. Stability analyses for the
deterministic model show that the final stable state is necessarily
racemic. Yet, the claim was made that reaction noise stabilizes the
homochiral states, making these the most probable outcome of the
system. According to this, in systems governed by stochastic
kinetics, i.e., before coinciding with the limit of deterministic
kinetics, the linear decay of the enantiomers to the compound A (Eq.
(\ref{direct})) is to play the role of mutual inhibition for
achieving SMSB, i.e., the homochiral states are supposed to result
without requiring additional ``non-linearities" nor even chiral
inhibition itself.

In this paper we analyze these claims in detail. As the need for
chiral inhibition in SMSB has been questioned \cite{Goldenfeld}, we
draw a careful distinction between linear racemization and
non-linear mutual inhibition in Sec \ref{sec:inhibs}. The
constraints dictated by detailed balance, often overlooked in the
modeling of biological homochirality, are discussed in Sec
\ref{sec:micro}. In Sec \ref{sec:results} we consider the influence,
for both well-mixed and small spatially coupled systems, that
reaction noise has on the stationary states of the chemical scheme
when detailed balance is properly accounted for. In Sec
\ref{sec:cycles} we address the problem of coupling chemical
reaction schemes to external energy sources for driving
\textit{unidirectional} cyclic reactions, a minimum requirement for
biological systems. Conclusions are drawn in Sec \ref{sec:disc}.
Details of the calculation of the probability distribution for the
enantiomeric excess are relegated to an Appendix.

\section{\label{sec:model} Chiral Inhibition and Microreversibility}

As the scheme in \cite{Goldenfeld} (i) dispenses with chiral
inhibition and (ii) overlooks the principle of detailed balance, we
consider these two concepts below, bringing in some closely related
reaction schemes for illustrative purposes.

\subsection{\label{sec:inhibs} Linear racemization and non-linear chiral inhibition}

The Frank model \cite{Frank} involves enantioselective autocatalysis
coupled to a reaction between the two enantiomers of
product/catalyst yielding an achiral addition product. The term
``mutual chiral inhibition" was coined for this reaction because it
represents the decrease of chiral compounds in a racemic ratio. This
leads to an increase in the value of the enantiomeric excess $ee$.
When this occurs faster than the reverse enantioselective
autocatalysis, namely when $g \equiv \frac{k_{-a}}{k_2} < 1$, a
cooperative effect drives the amplification of the $ee$ in the
enantioselective autocatalysis: Eq.(\ref{Frank1953}) or
Eq.(\ref{LESscheme}). As remarked above, in LES this mutual chiral
inhibition is manifested via the inverse non-enantioselective
autocatalysis, Eq.(\ref{limited}). Compared to Frank, in LES no
inert product P nor achiral heterodimer is formed, but instead the
recycling of one enantiomer back to the achiral precursor A plus the
mirror image enantiomer. This reverse reaction implies the
disappearance of only one stoichiometric part of the racemic mixture
(either D or L), but in the dynamics of the system there is a
non-linear dependence on the heterochiral interaction
$[L]\times[D]$, just as in the case of the mutual inhibition stage
of Frank-like systems. The importance for SMSB of coupling
enantioselective autocatalysis of first order (Eq.(\ref{Frank1953})
or Eq.(\ref{LESscheme})) with such inhibition stages [Eq.
(\ref{mutual}) or Eq.(\ref{limited})] is due to the fact that the
autocatalysis \textit{by itself} cannot yield SMSB. In lieu of these
inhibitions, there is no amplification of chirality, and in the best
of cases, (open flow systems, or systems with heterogeneous energy
distributions, etc.), the production of chiral matter can only
maintain the initial $ee$ value. The significance of the
heterochiral inhibition stage, when coupled to an enantioselective
autocatalysis of first order, is that the overall reaction network
is then able to lead to SMSB. This chiral state is a
non-thermodynamic one, but is the more stable state of the system.

The reaction of Eq. (\ref{direct}), has been reported
\cite{Goldenfeld} as being able to play the role of the needed
inhibition stage for SMSB in enantioselective autocatalysis of first
order, by appealing to the presence of reaction noise.  The
character of such an inhibition stage, however, is easily
appreciated by re-expressing it as follows:
\begin{eqnarray}
\label{direct1} L \autorightleftharpoons{$k_d$}{$k_{n}$}&A&
\autorightleftharpoons{$k_n$}{$k_{d}$} D,\\ \label{direct2}
\Rightarrow \,\, L &\autorightleftharpoons{$k_R$}{$k_R$}& D, \qquad
k_R = k_n \, k_d.
\end{eqnarray}
The reactions of Eqs.(\ref{direct},\ref{direct1}) are identical and
the direct transformation between enantiomers in Eq.(\ref{direct2})
represents an equivalent overall transformation, as far as L and D
are concerned. What actually differs between all three is the
temporal or spatial resolution or scale at which we can resolve them
into individual steps or else as overall, collective reactions. Thus
the first, Eq.(\ref{direct}), describes two individual reactions,
but because of the degenerate character of enantiomerism, the
existence of one reaction implies necessarily the existence of the
enantiomeric one. The second Eq. (\ref{direct1}) is identical to
(\ref{direct}), but it can also include a single reaction going
through a non-stable intermediate, and the third Eq. (\ref{direct2})
is a single reaction with no intermediate species made explicit. All
these reactions are the basic transformations representing the
chemical process known as ``racemization". The common effect of such
reactions determines, in the case of closed systems with homogeneous
energy distribution, that any initial $ee$ value must decrease
towards the unavoidable racemic mixture. Note that the three Eqs.
(\ref{direct}-\ref{direct2}) lead to the decrease the total chiral
matter, but the underlying dynamics does not involve a non-linear
dependence on the racemic composition \cite{AAS-review}. This
non-linearity is expressed in deterministic kinetics by the
dependence on the product of the concentrations of both enantiomers,
and in stochastic kinetics by the existence of a non-elastic
heterochiral collision between the enantiomers.

\subsection{\label{sec:micro} Microreversibility}

The rate constants in LES are constrained by the principle of
microreversibility \cite{RH}:
\begin{equation}\label{LESconstraint}
\frac{k_a}{k_{-a}} = \frac{k_e}{k_{-e}} = \frac{k_n}{k_{d}}.
\end{equation}
The LES model has had a controversial reputation in the past because
the constraints dictated by microreversibility or detailed balance,
have not always been correctly considered nor properly taken into
account \cite{RH,BM}. Using Eq.(\ref{LESconstraint}) one proves that
LES in a closed to mass flow system at uniform temperature cannot
lead to either temporary nor permanent chiral symmetry breaking: the
racemic state is the only stable outcome \cite{RH}. In order to
overcome these microreversibility constraints necessarily requires
extending the reaction model via coupling to external energy sources
and/or to external reagents. The new reactions or energy fluxes thus
introduced alter the overall set of (original) transformations and
can allow for the (partial) lifting of the original
microreversibility constraints. Thus for example, when the
enantioselective and the limited enantioselective autocatalyses are
individually \textit{localized} within regions of low and hot
temperatures, respectively, in a thermal gradient, mirror symmetry
can be broken permanently \cite{LES-temp,AAS-review}. Alternatively,
when the reverse reaction of the non-enantioselective autocatalysis
is driven by an external reagent, LES in a uniform temperature can
break mirror symmetry permanently \cite{LES-reagents,AAS-review}.
Both these modifications maintain LES far from equilibrium and also
lift the constraints Eq.(\ref{LESconstraint}) on some of the
reaction rates. Even so, this does not imply we can set any of the
inverse rate constants to zero (and we cannot: doing so would
violate Eq. (\ref{LESconstraint})).

Artifacts in mathematical modeling can and do arise when (i)
reactions are approximated by irreversible transformations and
especially when (ii) irreversible and reversible reactions are
combined \textit{together} in the same scheme \cite{BM}, as was done
in \cite{Goldenfeld}. Note that the original Frank scheme
Eqs(\ref{Frank1953},\ref{mutual}) involves only irreversible
reactions (understood as approximations), but here there is no
constraint dictated by microreversibility: in this case both rates
$k_1$ and $k_2$ are independent. The rates for the modified Frank
model with reversible autocatalysis and mutual inhibition: Eqs.
(\ref{LESscheme},\ref{mutual}), are also independent.

In contrast, the individual rate constants for \textit{reversible}
autocatalysis in concert with reversible non-catalytic production
must obey
\begin{equation}\label{constraint}
\frac{k_a}{k_{-a}} = \frac{k_n}{k_d} .
\end{equation}
We emphasize that, and as expressed by Wegscheider's rule
\cite{Wegscheider,Yablonsky}, the microreversibility constraint Eq.
(\ref{constraint}) requires us to include \textit{both} forward and
inverse chemical reactions in the autocatalysis Eq.
(\ref{LESscheme}), since direct production Eq. (\ref{direct}) is
taken to be reversible in \cite{Goldenfeld}. That is, one reaction
is reversible if and only if the other one is. Thus, we cannot set
$k_{-a} = 0$ in the presence of  Eq. (\ref{direct}). If however, we
insist on combining irreversible autocatalysis $k_{-a} = 0$ with
irreversible direct production $k_{d} = 0$ \cite{Lente}, then the
microreversibility constraint is satisfied consistently, and in the
most trivial way, since
\begin{eqnarray}
\frac{k_{-a}}{k_{a}} &=& \frac{k_d}{k_n},\\
\Rightarrow \, 0 &=& 0,
\end{eqnarray}
and we are free to vary the forward reaction rates $k_n$ and $k_a$
independently. But such an irreversible scheme \cite{Lente}
corresponds to reactions under strict kinetic control. On
asymptotically long time scales, the inverse reactions become
relevant and the closed mass system will necessarily racemize
\cite{CHMR}.

\section{\label{sec:results} Reaction noise: analytic results and simulations}

We consider the role of reaction noise in both well mixed and small
spatially coupled systems when detailed balance is taken into
account.

\subsection{\label{sec:wellmixed} Stochastic model: well-mixed system}

If we eliminate the reverse autocatalysis (set $k_{-a} = 0$) from
the following reactions
\begin{eqnarray}\label{autocata}
A + L  \autorightleftharpoons{$k_a$}{$k_{-a}$} L + L, &&\qquad A + D
\autorightleftharpoons{$k_a$}{$k_{-a}$} D + D, \\
\label{noncat} A
\autorightleftharpoons{$k_n$}{$k_{d}$} L, &&\qquad A
\autorightleftharpoons{$k_n$}{$k_{d}$} D.
\end{eqnarray}
we recover the scheme proposed in \cite{Goldenfeld}, which is itself
a minor variation of the model in Ref. \cite{Lente}.  When this
reverse step is overlooked, then the resultant reaction noise would
appear to stabilize the homochiral states, provided a certain
parameter $\alpha = V\frac{k_n}{k_a} \ll 1$.  Here we include the
obligatory reverse reaction as dictated by microreversibility, Eq.
(\ref{constraint}), and reconsider carefully the role of reaction
noise on the stationary states of the system. In passing we note
that the models considered up to this point are variations of either
the basic Frank paradigm or of the LES model, obtained by combining
some elements of the former with some elements of the latter, now
taking reversible steps, or instead taking irreversible steps, etc.

We thus approximate the scheme Eqs. (\ref{autocata}, \ref{noncat})
by means of a stochastic differential equation for the time
dependence of the enantiomeric excess $\theta = \frac{[D]-[L]}{[D] +
[L]}$. We consider a closed mass well-mixed system of volume $V$ and
total number of molecules $N$.  Taking the limit, $N \gg 1$, as in
\cite{Goldenfeld}, we arrive at the following equation for $\theta$
(see Appendix \ref{sec:stochastic}):
\begin{equation}\label{stochastic}
\frac{d \theta}{dt} = -\frac{k_{-a}}{2 + \frac{k_{-a}}{k_a}} \big(
\frac{N}{V} \big) \theta + \sqrt{\frac{k_{-a}}{2V}(1 - \theta^2)( 2
- \theta^2)} \,\,\eta(t),
\end{equation}
where $\eta(t)$ is Gaussian white noise with zero mean and unit
variance.

\begin{figure}[h]
\includegraphics[width=0.45\textwidth]{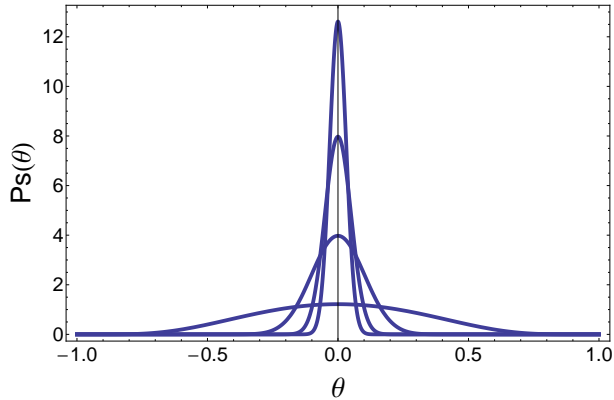}
\caption{\label{Fig1} Stationary probability distribution for the
chiral order parameter, Eq. (\ref{pstheta}). Different values of $b
= 10,100, 400, 1000$ correspond going from the broadest to the
narrowest distribution. $P_s(|\theta| = 1) = 0$ for homochiral
states. }
\end{figure}

\begin{figure}
\includegraphics[width=0.30\textwidth]{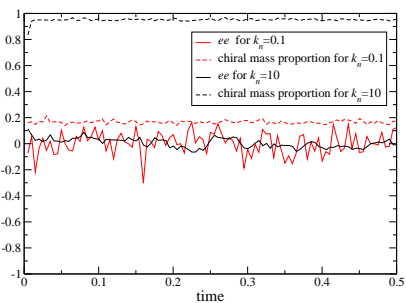}
\caption{\label{time-series} Temporal series for the enantiomeric
excess $ee$ and the chiral mass proportion obtained from Gillespie
simulations for different values of $k_n$ (see legend). After a very
brief transient, the curves fluctuate about the racemic state. The
parameters are: $k_a=k_d=1$ (and hence $k_{-a}=1/k_n$), the number
of molecules is 1000 (initial condition is 10 L, 10 D, 980 A). The
$ee$ is defined as $([L]-[D])/([L]+[D])$ and the chiral mass
proportion as $([L]+[D])/([L]+[D]+[A])$.}
\end{figure}

The normalized stationary distribution of Eq. (\ref{stochastic}) is
given by
\begin{equation}\label{pstheta}
P_s(\theta) =  \frac{2^{1+b} \Gamma(b + \frac{1}{2})}{\sqrt{\pi}
\Gamma(b) F(\frac{1}{2},1+b,\frac{1}{2}+b; \frac{1}{2})}\,\frac{(1 -
\theta^2)^{b-1}}{(2 - \theta^2)^{b+1}}, \qquad {\rm with} \qquad b =
\frac{N}{1 + \frac{k_{-a}}{2 k_a}}.
\end{equation}
We plot $P_s(\theta)$ for various values of $b$ in Fig \ref{Fig1}.
The distribution $P_s(\theta)$ is \textit{always} peaked around the
racemic state $\theta = 0$ since the parameter $b \gg 1$. As the
total number of molecules $N$ increases, the distribution becomes
ever more sharply peaked around $\theta = 0$. In particular, the
probability for homochiral states $|\theta| = 1$ is strictly zero.
The deterministic part of Eq.(\ref{stochastic}) has one fixed point
at the racemic state $\theta=0$, in accord with stability analyses
for the deterministic kinetic rate equations. The amplitude of the
noise is maximum for the racemic state, and vanishes at the
homochiral states. Nevertheless, we cannot arrange for the noise
amplitude to be larger than that of the deterministic term, since $b
\gg 1$; see Eq. (\ref{pstheta}). This means that the racemic state
is stable in the presence of reaction noise, and is surrounded by
Gaussian fluctuations that become negligible for increasing total
number $N$ of molecules in the system, see Fig \ref{Fig1} and Fig.
\ref{time-series}.
\vspace{1cm}
As shown in Fig. \ref{time-series}, stochastic simulations of the
scheme Eqs. (\ref{autocata},\ref{noncat}) using the Gillespie
algorithm \cite{Gillespie} reveal that the magnitude of the
fluctuations about the racemic composition depend on the rate $k_n$.
Thus we observe that the reaction noise is somewhat more erratic for
$k_n = 0.1$ in comparison with the smoother fluctuations that result
when $k_n= 10$. Note moreover the dependence of the total
\textit{chiral mass proportion}, defined as
$([L]+[D])/([L]+[D]+[A])$: the fraction of total system mass which
is chiral. Increased non-catalytic production leads to a greater
proportion of chiral matter. The other rates were set to $k_a=k_d=1$
as in \cite{Goldenfeld}, and we include $k_{-a}=1/k_n$ as dictated
by microreversibility. This implies that smaller $k_n$ thus leads to
a greater recycling of the enantiomers back to achiral precursor via
reverse autocatalysis, leading to smaller net chiral matter than
when $k_n$ is large.

\begin{figure}[h]
\includegraphics[width=0.35\textwidth]{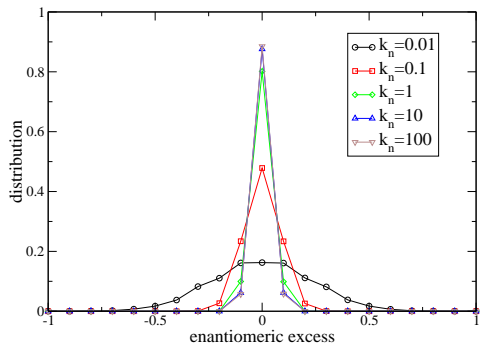}
\caption{\label{R1PP} Distribution of the enantiomeric excess $ee$
obtained from Gillespie simulations for different values of $k_n$
(see legend). After a brief initial transient, the distributions
center about the racemic state. The parameters are: $V=1$,
$k_a=k_d=1$ (and hence $k_{-a}=1/k_n$), the number of molecules is
1000 (initial condition is 10 L, 10 D, 980 A). We obtain the $ee =
([L]-[D])/([L]+[D])$ at $T=5$. We perform $R=100000$ realizations.
Binning is in intervals of $0.1$ in the enantiomeric excess. }
\end{figure}

The racemizing tendency of the forward rate of non-catalytic
production can also be appreciated in Fig \ref{R1PP} which shows the
distribution in the enantiomeric excesses for different values of
$k_n$. The greater the $k_n$, the more sharply peaked is the
distribution about the racemic outcome; compare to Fig \ref{Fig1}.

%

\subsection{\label{sec:2patch} Stochastic model: spatial coupling}

\begin{figure}
\includegraphics[width=0.35\textwidth]{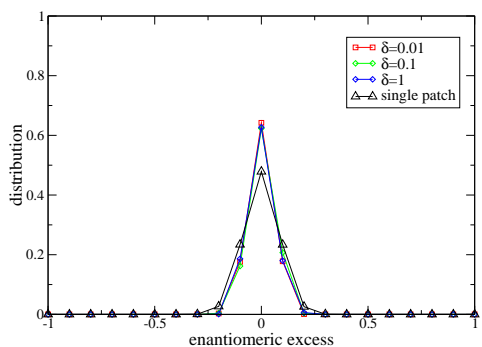}
\caption{\label{two-patch} The distribution of the total
enantiomeric excess obtained from Gillespie simulations of a
two-patch model for $k_n=0.1$ and different values of $\delta$ (see
legend). Asymptotically, the distributions center around the racemic
state. The parameters are: $k_a=k_d=1$ (and hence $k_{-a}=10$),
$V=1$, the initial number of molecules per patch is 1000 (10 L, 10
D, 980 A) and we obtain the total enantiomeric excess $([L_1]-[D_1]+
[L_2]-[D_2])/([L_1]+[D_1]+[L_2]+[D_2])$ at $T=20$ and perform
$R=1000$ realizations. Binning is in intervals of $0.1$ in
enantiomeric excess. Spatial coupling applies to L, D, and A. For
comparison, we show the distribution for a single uncoupled patch
with $k_n=0.1$ (100000 realizations).}
\end{figure}

The reaction scheme can be generalized and studied by spatially
coupling a number of well-mixed systems.  In the manner of
\cite{Goldenfeld}, space can be discretized into a set of $M$
patches of volume $V$, with the patches indexed by $i$. Here we
consider such ``spatial coupling" of $M=2$ well-mixed patches.
Within each such patch the reactions Eqs.
(\ref{autocata},\ref{noncat}) take place, with identical reaction
rates for each patch, while all the molecules $A_i,L_i,D_i$ can
diffuse from one patch to the other, with a common spatial coupling
constant $\delta$ (or, intra-patch ``diffusion"):
\begin{equation}
A_i  \stackrel{\delta}{\rightleftharpoons} A_j, \qquad D_i
\stackrel{\delta}{\rightleftharpoons} D_j, \qquad L_i
\stackrel{\delta}{\rightleftharpoons}L_j, \qquad i\neq j\in (1,2).
\end{equation}
Stochastic simulations (Gillespie algorithm) indicate that each
patch racemizes independently. The curves for the two patch system
do not depend on the spatial coupling constant $\delta$, see Fig.
(\ref{two-patch}).

\vspace{2.5cm}
%

\section{\label{sec:cycles} Unidirectional irreversible cycles}

In order to maintain an ``irreversible self-replication", the system
in \cite{Goldenfeld} would have to be driven by an external source
of energy to maintain the steady state of the system far from
equilibrium. This is a necessary, but not sufficient, condition for
achieving SMSB. Additional effort is required to include explicitly
such a hypothetical source as an integral part of the overall model,
and then to demonstrate its feasibility. Indeed, perhaps the most
challenging aspect of any proposal for modeling biological
homochirality at the molecular level is not so much in the design of
the intermediate reaction scheme itself, but rather in defining the
very nature of the essential external energy source and its coupling
to the reactions that comprise the intermediate chemical system.

The sought-after ``irreversible self-replication" corresponds to a
unidirectional cyclic reaction. But setting up such a cyclic
reaction, by no means implies that the elementary autocatalytic
reaction be ``irreversible", i.e., by simply putting $k_{-a} = 0$.
The true cyclic behavior requires maintaining a steady
unidirectional flow of matter in the system. This unidirectional
cycling behavior is a general property of steady states maintained
by an energy flux. A brief review of the meaning of \textit{cycle},
using Onsager's original ``triangle reaction" as an example,
\cite{Onsager} is warranted. Consider a closed mass system held at
constant temperature, containing species A,B and C reacting
according to the scheme:
\begin{equation}\label{Onsager}
A \autorightleftharpoons{$k_1$}{$k_2$} B
\autorightleftharpoons{$k_3$}{$k_4$} C
\autorightleftharpoons{$k_5$}{$k_6$} A.
\end{equation}
At equilibrium there is no net flow around system since the forward
and inverse reactions are in detailed balance( $[.]_{eq}$ denotes
the equilibrium concentration):
\begin{equation}
k_1[A]_{eq} - k_2[B]_{eq} = k_3[B]_{eq} - k_4[C]_{eq} = k_5[C]_{eq}
- k_6[A]_{eq} = 0.
\end{equation}
The principle of microreversibility implies the following
Wegscheider condition for the rate constants:
\begin{equation}\label{microreverse}
k_1 k_3 k_5 = k_2 k_4 k_6.
\end{equation}
An external energy source could drive a net flow of material around
the system, brought about by a flow of energy from a high potential
source to a low potential sink, passing through the closed mass
intermediate system of Eq.(\ref{Onsager}).  The steady state
condition would then require that
\begin{equation}\label{steadyflow}
k_1[A] - k_2[B] = k_3[B] - k_4[C] = k_5[C] - k_6[A] = {\cal F} > 0,
\end{equation}
where ${\cal F}$ is the flow, the rate at which material is cycling
around the system. This flow of material around a closed reaction
loop is what is meant by a cycle \cite{Morowitz}.  Note that this
cycle is irreversible: the net matter flow is unidirectional. We
emphasize two important points: (i) Eq.(\ref{steadyflow}) involves
the non-equilibrium concentrations of the species involved, and (ii)
the unidirectional matter flow ${\cal F}$ depends on \textit{all}
the forward and reverse reaction rates $k_i$. The cycle is
\textit{not} established by simply putting the reverse rates to zero
(this is prohibited by the Wegscheider condition Eq.
(\ref{microreverse})), but rather from an energy flow that traverses
the closed mass system. By way of example, Morowitz offers a kinetic
model for unidirectional cycles in Onsager's network using
photochemical reactions \cite{Morowitz}.

\begin{figure}[h]
\includegraphics[width=0.50\textwidth]{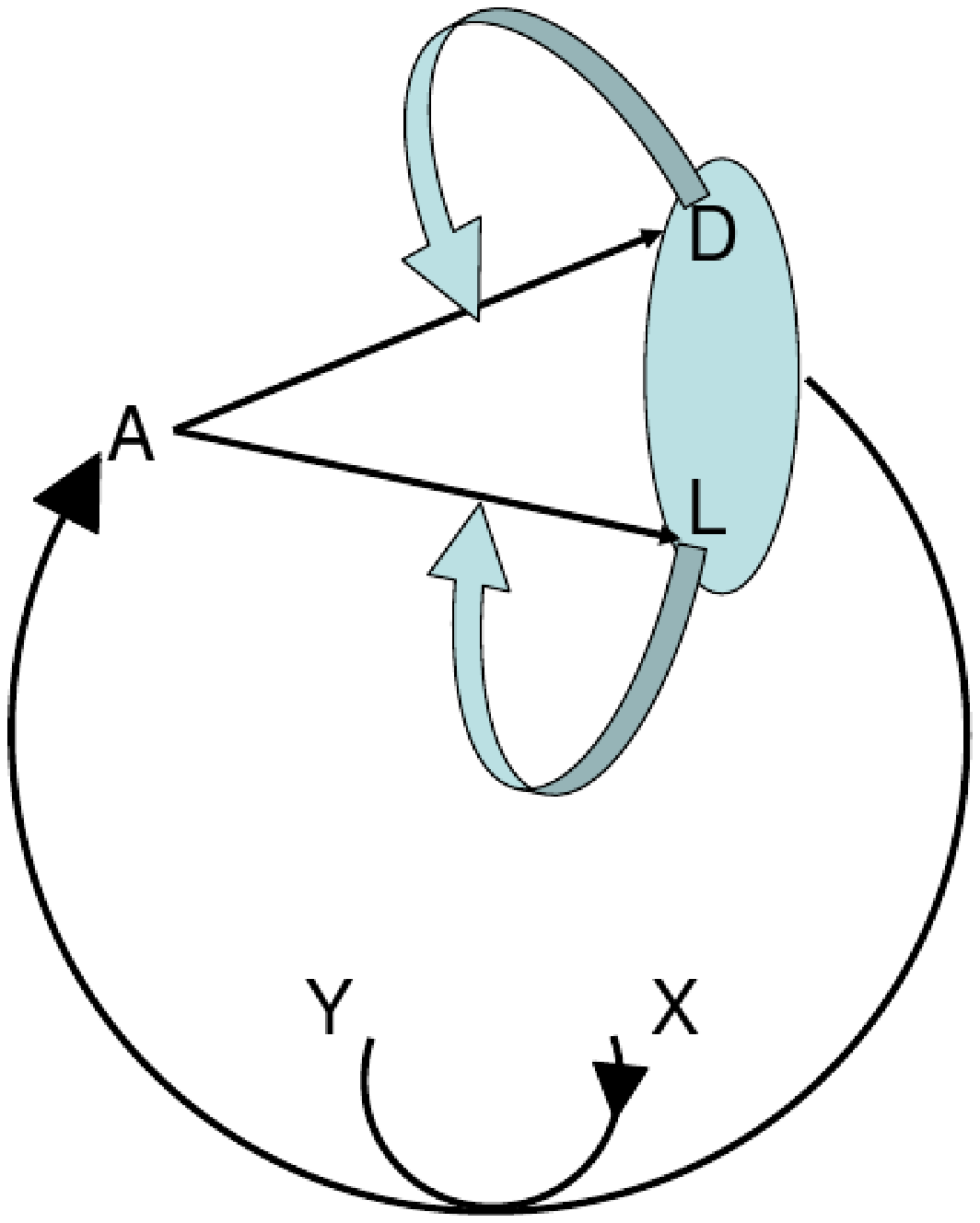}
\caption{\label{Cycle} The unidirectional cyclic network discussed
in the text is composed of the \textit{micro-reversible} reactions
Eqs.(\ref{drivecata}, \ref{noncat2}) in a uniform temperature
distribution with Y and X matter exchange with the environment. A is
an achiral compound and D and L the enantiomer pair of a chiral
compound. The cycle is driven by the external reagents.}
\end{figure}

In Sec \ref{sec:results}, we demonstrated that the reaction scheme
Eqs. (\ref{autocata},\ref{noncat}) when obeying Eq.
(\ref{constraint}), does not lead to SMSB, regardless of the
inclusion of reaction noise. We can legitimately circumvent this
latter constraint by going to an out-of-equilibrium scenario. To
assess whether irreversible cycling can lead to SMSB, we consider
this scheme in a uniform temperature distribution driven by a
constant concentration of external reactants, X and Y. See  Eq.
(\ref{drivecata}), an open system with X and Y matter exchange with
the surroundings, and depicted in Fig \ref{Cycle}.  The resultant
reaction network is cyclic one with permanent consumption and
production of Y and X (or of X and Y, depending on the flow
direction in the cycle):
\begin{eqnarray}\label{drivecata}
A + L + Y\autorightleftharpoons{$k_2$}{$k_{-2}$} L + L + X, &&\qquad
A + D + Y
\autorightleftharpoons{$k_2$}{$k_{-2}$} D + D + X, \\
\label{noncat2} A \autorightleftharpoons{$k_n$}{$k_{d}$} L, &&\qquad
A \autorightleftharpoons{$k_n$}{$k_{d}$} D.
\end{eqnarray}
It is straightforward to carry out a dynamic stability analysis for
this system. The presence of external constant concentration
reactants now lifts the constraint Eq. (\ref{constraint}), and the
far-from-equilibrium reaction model depends on the \textit{two}
independent parameters:
\begin{equation}
u = \frac{k_{d}}{k_n}, \qquad g = \frac{k_{-2}[X]}{k_2[Y]}.
\end{equation}
Because the reactants are external to the intermediate system, we
can control the matter flow, e.g., in the forward sense, see Fig.
\ref{Cycle}:
\begin{eqnarray}
k_2[A][Y] - k_{-2}[L][X] &=& {\cal F}_l > 0,\\
k_2[A][Y] - k_{-2}[D][X] &=& {\cal F}_d > 0.\\
\end{eqnarray}
The matter flow ${\cal F}_{tot}$ is partitioned ${\cal F}_l + {\cal
F}_d = {\cal F}_{tot}$, among the $L$ or $D$ autocatalytic branch of
the reaction network, respectively. Nevertheless, even with the
enantioselective autocatalyses driven \textit{unidirectionally} in
this way (${\cal F}_l > 0, {\cal F}_d > 0$), a stability analysis
proves that the racemic state is the only stable outcome, and for
all $u,g > 0$.

By marked contrast, whereas detailed balance implies the racemic
outcome for the LES model in a closed-mass system at uniform
temperature, driving LES by external reagents can lead to SMSB
\cite{LES-reagents}. The crucial fundamental difference between
Eqs(\ref{drivecata},\ref{noncat2}) and LES driven by external
reagents \cite{LES-reagents} is in the inverse limited
enantioselective catalytic step Eq.(\ref{limited}), which implies a
\textit{non-linear chiral inhibition} between the two enantiomers
\cite{driveFrank}.

\section{\label{sec:disc} Concluding Remarks}

The reaction models analyzed in this paper have served as useful
vehicles for examining a number of basic issues relevant for framing
proposals, coherent with fundamental chemical and physical
principles,  aimed at modeling biological homochirality at the
molecular level. We summarize our main findings here. The following
points are, to some extent, inextricably interrelated.

\begin{itemize}

  \item \textsf{Detailed balance}. Thermodynamics dictates that
  the enantioselective autocatalysis and direct production/decay of the enantiomers must
  have identical ratios of the forward and reverse reaction rate constants,
regardless of whether the system is in equilibrium or far from it.
Once detailed balance is accounted for correctly, we have proven, by
employing standard methods (stochastic differential equations, the
Fokker-Planck equation) and numerical simulations, that the
resultant model including reaction noise can never break chiral
symmetry. On the contrary, the final stable outcome is always the
racemic state. And, this holds whether the system is well-mixed or
coupled spatially.

  \item \textsf{Reaction noise}. We have proven that the presence of
reaction noise does not lead to any new final stable state not
already in accord with the stability analysis of the deterministic
model. Note: the regime where stochastic kinetics is expected to be
important corresponds to the case of small volumes and
\textit{small} numbers $N$ of molecules. In deriving our stationary
probability distribution, we take the limit $N \gg 1$, as do the
authors of Ref. \cite{Goldenfeld}. In this limit, reaction noise has
only a minor quantitative, but not qualitative, effect.

  \item \textsf{Chiral inhibition}.  We have argued that the linear decay reaction of
Eq.(\ref{direct}) cannot act as a mutual inhibition stage for SMSB
when coupled to a first-order enantioselective autocatalysis. This
strongly suggests, that a chiral inhibition reaction, or a set of
coupled reactions generating a chiral inhibition dynamics, must have
a \textit{non-linear} chiral dependence \cite{Ecology}, as is the
case in both the LES and Frank models.

  \item \textsf{Irreversible cycles}. We have shown explicitly how to establish a unidirectional net flow of matter in
the reversible autocatalytic
  reaction Eq.(\ref{drivecata}), by coupling it to external reagents.  In spite of this,
  a stability analysis proves that
   the manifestly out-of-equilibrium model Eqs.(\ref{drivecata},\ref{noncat2}) leads inexorably to the racemic
   state. And this result is intimately related to the absence of
   chiral inhibition.

\end{itemize}

There is a widespread and active research effort devoted to
understanding the origins of biological homochirality that crosses
the traditional boundaries between physics, chemistry and biology.
The fundamental concepts treated here deserve careful consideration
in scenarios for candidate reaction schemes proposed as models for
the emergence of biological homochirality.

\begin{acknowledgments}
The research of JMR and DH is supported in part by the Coordinated
Project CTQ2013-47401-C2-1/2-P (MINECO). MS, JMR and DH form part of
the COST Action CM1304 on ``Emergence and Evolution of Complex
Chemical Systems".
\end{acknowledgments}

\appendix

\section{\label{sec:stochastic} Probability distribution for the enantiomeric excess $\theta$}

We cast the fully reversible kinetic scheme defined by Eqs.
(\ref{autocata},\ref{noncat}) in terms of stochastic differential
equations to quantify the role played by internal reaction noise.
The mapping of chemical reactions to master equations and then on to
Fokker-Planck (FP) equations is an established technique
\cite{Gardiner,Risken}, as is the correspondence of FP with
stochastic differential equations. Defining the state vector
$\vec{x} = (x_1,x_2,x_3) \equiv (a,d,l)$ where $a,d,l$ denote the
time-dependent concentrations of molecules A,D and L, respectively,
we find that our scheme may be approximated by the stochastic
differential equation (defined in the Ito sense) \cite{SM-Goldy}:
\begin{equation}\label{Ito}
\frac{d \vec{x}}{d t} = \vec{H}(\vec{x}) +
\bm{G}(\vec{x})\vec{\eta}(t),
\end{equation}

where
\begin{equation}\label{H}
\vec{H} = \left(
              \begin{array}{c}
                k_{-a}(d^2+l^2) -a(2k_n + k_a(d+l))+ k_d(d+l) \\
                -k_{-a}d^2 + a(k_n + k_a d) -k_d d\\
                -k_{-a}l^2 + a(k_n + k_a l) -k_d l \\
              \end{array}
            \right),
\end{equation}
\begin{equation}\label{G}
\bm{G} = \frac{1}{\sqrt{V}} \left(
           \begin{array}{cc}
             \sqrt{k_{-a}d^2 + a(k_a d + k_n) + k_d d} & \sqrt{k_{-a}l^2 + a(k_a l + k_n) + k_d l} \\
             -\sqrt{k_{-a}d^2 + a(k_a d + k_n) + k_d d} &  0\\
             0 & -\sqrt{k_{-a}l^2 + a(k_a l + k_n) + k_d l} \\
           \end{array}
         \right),
\end{equation}
an the $\eta_j$ $(j=1,2)$ are Gaussian white noises with zero mean
and correlation, $< \eta_i(t) \eta_j(t') > = \delta_{ij}
\delta(t-t').$  $V$ is the system volume. The rate of inverse
autocatalysis is \textit{not} an independent variable, but obeys the
constraint:
\begin{equation}
k_{-a} = k_a \frac{k_d}{k_n}.
\end{equation}
The number of chemical degrees of freedom $\vec{x}$ can be
effectively reduced from three to one \cite{SM-Goldy}. This is so
because firstly, the total number of molecules is conserved by our
reaction scheme, thus so is the total concentration $n = a + d + l$.
Secondly, the total chiral matter $\chi = d + l$ is a \textit{fast
degree of freedom} relative to the enantiomeric excess $\theta$
\cite{HZpre}. Simulations of the fully reversible scheme Eqs.
(\ref{autocata},\ref{noncat}) using the Gillespie algorithm
\cite{Gillespie} confirm that $\chi$ approaches a stable fixed point
value surrounded by small Gaussian fluctuations (see, e.g., Fig
\ref{time-series}). We therefore substitute $\chi(t) \rightarrow
\chi^*$ into the equation for $\theta(t)$ derived below. We thus
carry out the change of variables on Eq.(\ref{Ito}):
\begin{eqnarray}
(a,d,l) \rightarrow (n,\chi, \theta) = (a + d + l, d+l,
(d-l)/(d+l)),
\end{eqnarray}
employing Ito's formula \cite{Gardiner}:
\begin{equation}\label{Ito-change}
df(\vec{x}) = [\sum_i H_i(\vec{x}) \partial_i f(\vec{x}) +
\frac{1}{2}\sum_{i,j} [\bm{G}\bm{G}^T]_{ij} \partial_i \partial_j
f(\vec{x})] dt + \sum_{ij} \bm{G}(\vec{x})_{ij} \partial_i
f(\vec{x}) dW_j(t).
\end{equation}
From Eq. (\ref{Ito-change}) it is straightforward to demonstrate
that $\frac{dn}{dt} \equiv 0$ is identically zero, as it must be.
From $\frac{d \chi}{dt} = 0$ we solve for the fixed point $\chi^*$:
\begin{equation}\label{chi-star}
\chi^*(\bar{\theta}) = \frac{k_a n -2k_n -k_d + \sqrt{(k_a n -2k_n
-k_d)^2 + 8n k_n[k_a + \frac{1}{2}(1 + \bar{\theta}^2)k_{-a}]}}{2k_a
+ (1 + \bar{\theta}^2)k_{-a}},
\end{equation}
Since $\chi \geq 0$, we take the positive root. Note the total
chiral matter $\chi^*(\bar{\theta})$ depends \textit{weakly} on the
most probable stationary value $0 \leq \bar{\theta}^2 \leq 1$ for
the chiral order parameter. The most probable value of
$\bar{\theta}$ is determined from the stochastic differential
equation for $\theta(t)$. We prove below that
\textit{self-consistency} requires taking $\bar{\theta} = 0$ in Eq.
(\ref{chi-star}).

We derive the stochastic equation obeyed by $\theta(t)$ and
substitute $\chi^*(0)$ into this equation. We express the result in
terms of the total number of molecules $N = V n$ and for $N \gg 1$.
The enantiomeric excess or chiral order parameter $\theta$ obeys the
equation
\begin{equation}\label{theta-eq}
\frac{d \theta}{dt} = -\frac{k_{-a}}{2 + \frac{k_{-a}}{k_a}} \big(
\frac{N}{V} \big) \theta + \sqrt{\frac{k_{-a}}{2V}(1 - \theta^2)( 2
- \theta^2)} \,\,\eta(t),
\end{equation}
where $\eta(t)$ is Gaussian white noise with zero mean and unit
variance.

From the Fokker-Planck equation corresponding to Eq (\ref{theta-eq})
we readily solve for the steady state probability distribution
$P_s(\theta)$ for $\theta$ \cite{Risken}.  We find
\begin{equation}
P_s(\theta) = {\cal N} \frac{(1 - \theta^2)^{b-1}}{(2 -
\theta^2)^{b+1}}, \qquad {\rm with} \qquad b = \frac{N}{1 +
\frac{k_{-a}}{2 k_a}},
\end{equation}
and the normalization constant
\begin{equation}\label{normalization}
{\cal N} = \Big( \int_{-1}^{1} d\theta \,\frac{(1 -
\theta^2)^{b-1}}{(2 - \theta^2)^{b+1}} \Big)^{-1} = \frac{2^{1+b}
\Gamma(b + \frac{1}{2})}{\sqrt{\pi} \Gamma(b)
F(\frac{1}{2},1+b,\frac{1}{2}+b; \frac{1}{2})},
\end{equation}
where $F$ is the hypergeometric function \cite{Hochstadt}.

From $P_s$ we conclude (see Fig \ref{Fig1}) that the most probable
value for the chiral order parameter is $\bar \theta = 0$,
corresponding to the racemic state, thus establishing the
self-consistency of employing this value in Eq. (\ref{chi-star}).


\begin{thebibliography}{99}
\bibitem{smsb} Systems
capable of SMSB yield large net chiral excesses, mostly in the
vicinity of homochirality, as common non-thermodynamic states
\cite{AAS-review}. For example, in closed systems with relatively
high exergonic transformations, prolonged chiral excursions i.e.,
kinetically controlled chiral outcomes, are possible. Even final
stationary chiral states are possible in either open or in closed
systems unable to come into thermodynamic equilibrium with their
surroundings; see, for example, Refs.
\cite{LES-temp,LES-reagents,CHMR}.
\bibitem{amplify} This amplification does not
correspond to the term as used in chemical asymmetric synthesis,
because in the absence of any external chiral force or polarization,
the racemic mixture is a metastable, but not the most stable, state
of the system.
\end{thebibliography}
\end{document}